\documentclass[twocolumn]{aastex631}
\usepackage{graphicx}
\usepackage{booktabs}
\usepackage{multirow}
\usepackage{gensymb}
\usepackage{bm}

\newcommand\lsim{\mathrel{\rlap{\lower4pt\hbox{\hskip1pt$\sim$}}
\raise1pt\hbox{$<$}}}
\newcommand\gsim{\mathrel{\rlap{\lower4pt\hbox{\hskip1pt$\sim$}}
\raise1pt\hbox{$>$}}}
\begin{document}

\title{

Natal Kicks from the Galactic Center and Implications on their Environment and the Roman Space Telescope

}
\correspondingauthor{Carlos Jurado}
\email{carx207@g.ucla.edu}
\author[0009-0009-7568-8851]{Carlos Jurado}
\affiliation{Department of Physics and Astronomy, University of California, Los Angeles, CA 90095, USA}

\author[0000-0002-9802-9279]{Smadar Naoz}
\affiliation{Department of Physics and Astronomy, University of California, Los Angeles, CA 90095, USA}
\affiliation{Mani L. Bhaumik Institute for Theoretical Physics, Department of Physics and Astronomy, UCLA, Los Angeles, CA 90095, USA}

\author[0000-0002-6406-1924]{Casey Y. Lam}
\affiliation{Department of Astronomy, University of California, Berkeley, CA 94720, USA}
\affiliation{Observatories of the Carnegie Institution for Science, Pasadena, CA 91101, USA}

\author[0000-0003-0992-0033]{Bao-Minh Hoang}
\affiliation{Department of Physics and Astronomy, University of California, Los Angeles, CA 90095, USA}
\begin{abstract}
Most galaxies, including the Milky Way, harbor a central supermassive black hole (SMBH) weighing millions to billions of solar masses. Surrounding these SMBHs are dense regions of stars and stellar remnants, such as neutron stars and black holes. Neutron stars and possibly black holes receive large natal kicks at birth on the order of hundreds of km~s$^{-1}$.  The natal kicks that occur in the vicinity of an SMBH may redistribute the orbital configuration of the compact objects and alter their underlying density distribution. We model the effects of natal kicks on a Galactic Center (GC) population of massive stars and stellar binaries with different initial density distributions. Using observational constraints from stellar orbits near the GC, we place an upper limit on the steepness of the initial stellar profile and find it to be core-like. In addition, we predict that $30-70 \%$ of compact objects become unbound from the SMBH due to their kicks and will migrate throughout the galaxy. Different black hole kick prescriptions lead to distinct spatial and kinematic distributions. We suggest that the Roman Space Telescope may be able to distinguish between these distributions and thus be able to differentiate natal kick mechanisms.

\end{abstract}
\section{Introduction} \label{sec:intro}
Nuclear star clusters (NSCs) are the dense regions consisting of stars and stellar remnants near the centers of most galaxies, including our Milky Way. Most NSCs surround a central supermassive black hole (SMBH) with a mass between $10^6 - 10^9 \: {M}_\odot$ \citep[e.g.,][]{Kormendy+95,Ghez+00,Ghez+08,Ferrarese+05,Gillessen+09,Kormendy+Ho}.  Due to its proximity, our Galactic Center (GC) can serve as a unique place to investigate the conditions likely to occur at other galactic nuclei.


While the star formation process in the vicinity of an SMBH still remains a mystery, in particular with respect to the prevalence of binary formation, some studies indicate similarities to the field, where most massive stars (OBA spectral type) reside in a binary or higher order configuration \citep[e.g.,][]{Raghavan+10,Moe+17,Sana+12}. Specifically, there are already three confirmed eclipsing binaries in the inner $\simeq 0.2$~pc of the GC \citep[e.g.,][]{Ott+99,Martins+06,Rafelski+07,Pfuhl+14}, with possibly even more candidates \citep[e.g.,][]{Jia+19,Gautam+19}. Observations of the inner $0.02$~pc find a dearth of young few million year old binaries, consistent with dynamical interactions \citep{Chu+23} and suggesting a binary fraction close to $100$\% at birth for massive S-cluster stars \citep[e.g.,][]{Stephan+16}. Furthermore, X-ray observations have detected a large number of X-ray sources, implying a population of X-ray binaries or cataclysmic variables \citep[e.g.,][]{Muno+05,Hailey+18}. 

On the theoretical side, \citet{Stephan+16} suggested that as many as $70\%$ of binaries survive after a few million years of dynamical evolution at the GC. The dynamical interaction includes both frequent flybys from single passing stars that tend to unbind the binary \citep[known as evaporation process;][]{BT,Rose+20}, as well as interaction with the SMBH via the Eccentric Kozai Lidov mechanism \citep[EKL;][]{Kozai,Lidov,Naoz16}. Further, \citet{Naoz+18} suggested that the existence of binaries may  explain the peculiar properties of the stellar disk in the GC \citep{Yelda+14}. Moreover, merging binaries were suggested to form the G2-like object population \citep[e.g.,][]{Witzel+14,Witzel+17,Stephan+16,Stephan+19,Ciurlo+20}.  

The evolution of massive binaries in the GC is affected by natal kicks that neutron stars (NSs), and possibly black holes (BHs), receive at birth \citep{Fragione+19, Lu+19, Hoang+22}. 
Observations of pulsar motion have revealed that neutron stars receive significantly large kick velocities on the order of hundreds of km~$s^{-1}$ \citep[e.g.,][]{Hansen+Phinney97, Lorimer+Bailes97, Cordes+Chernoff98, Fryer+99, Hobbs+04}. It has been demonstrated that natal kicks can account for the misalignment between the orbital angular momentum and spin axes observed in pulsar binaries \citep[][]{Lai+95, Kalogera96, Kaspi+96, Kalogera+98, Kalogera00}. Studies have suggested that hypervelocity stars (HVSs) \citep[e.g.,][]{Zubovas+13, Bortolas+17, Fragione+17, Lu+19}, as well as extreme mass ratio inspirals (EMRIs) can be produced as a result of natal kicks disrupting massive binaries in the GC \citep[e.g.,][]{Bortolas+19, Lu+19, Hoang+22}.

It is currently debated as to what the underlying stellar and stellar remnant distribution around SMBHs at the center of galaxies is. Theoretical arguments of a dynamically relaxed population yield, $\rho(r) \propto r^{-\alpha},$ with $\alpha = 3/2 - 11/4$ \citep{Bahcall+77, Alexander+09, Keshet+09}. However, detailed measurements of the stars in our GC suggest a shallower distribution of $\alpha$ = 1.1 - 1.4 \citep{Schodel+18, Gallego-Can+18}. The distribution of compact objects at the GC, also known as the ``dark cusp'', has important implications for the dynamics in the vicinity of an SMBH. In particular, the compact object distribution strongly affects the rate of gravitational wave events, tidal disruption events, and the fraction of long-lived binaries in the GC \citep[][]{Amaro+07,Alexander11, Pfuhl+14, Rose+20}.

In this work, we study the evolution of binary stars orbiting the Galactic Center's SMBH and the resultant distribution of NSs and BHs. In Section \ref{Sec:methods}, we describe the methodology to form single and binary BH and NS systems from massive stellar binaries, as well as the different natal kick prescriptions. In Section \ref{Sec:PredictionsATGC}, we show that varying the initial stellar distribution steepens the post-kick compact object distribution, and that observations of the unseen mass in the Galactic Center allow us to constrain the initial stellar density profile. We also find that numerous high-energy events will be produced in this environment. In Section \ref{Sec:PredictionForRoman}, we study the spatial and velocity distribution of compact objects near the Galactic Center, and suggest that the Roman Space Telescope may be able to distinguish between different kick prescriptions. We close with discussion and conclusions in Section \ref{Sec:DiscAndConclusion}.

\section{Methodology}
\label{Sec:methods}

\begin{figure*}[t]
    \centering
    \includegraphics[width=18cm]{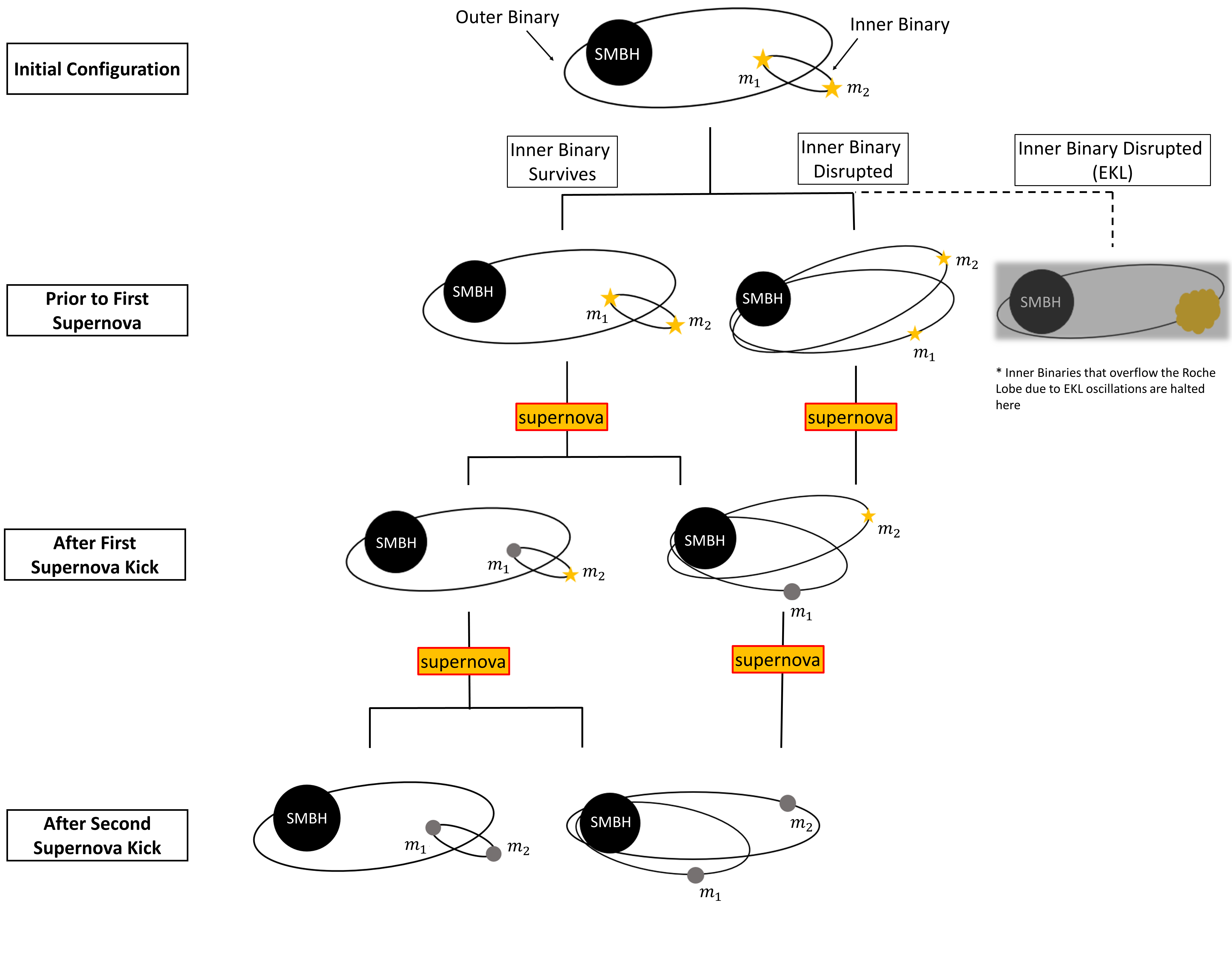}
    \caption{A simplified diagram illustrating our system setup and subsequent evolution. Note that this diagram does not depict every single outcome of natal kicks in a binary nor compact objects unbound from the SMBH due to the natal kicks.}
    \label{fig:Possible_Configs}
\end{figure*}

\begin{figure*}[t]
    \centering
    \includegraphics[width=18cm]{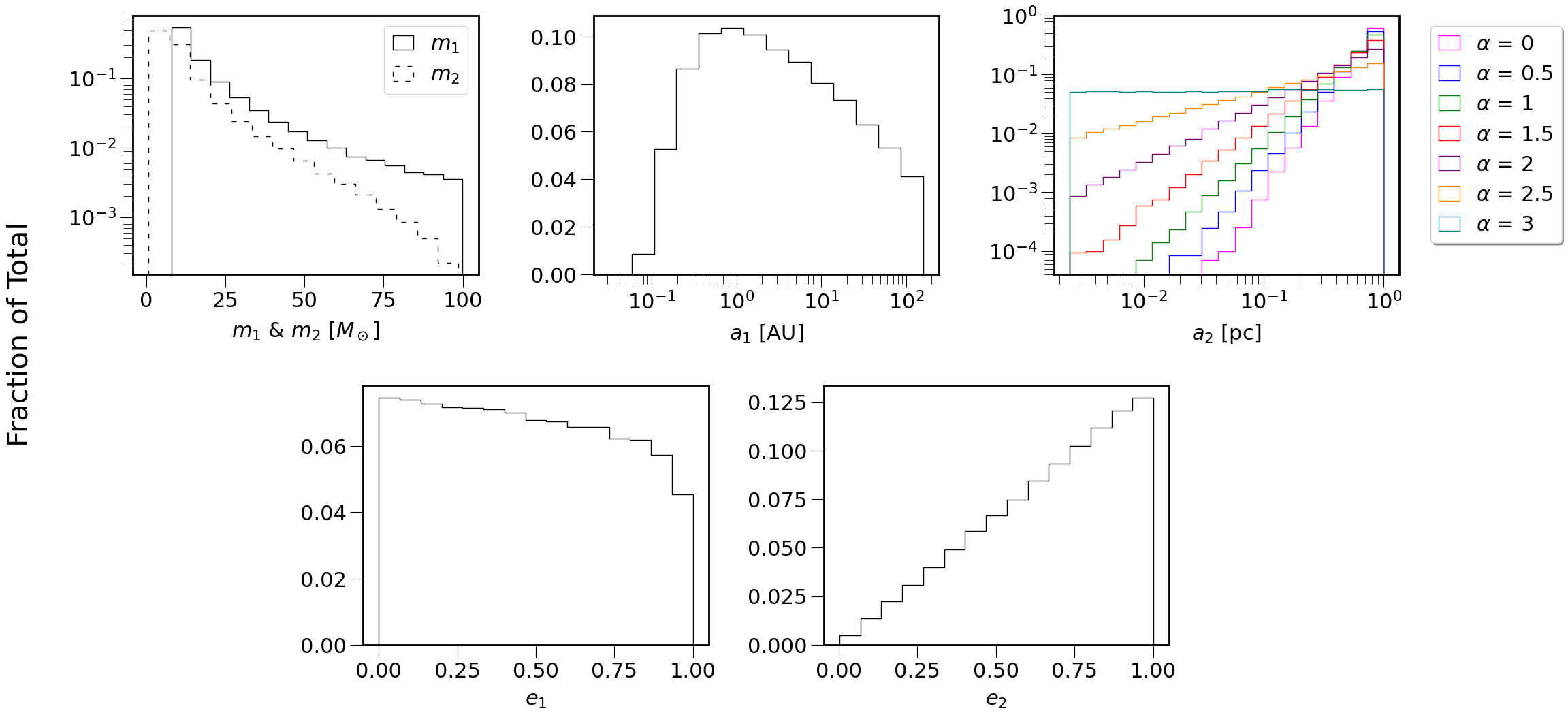}
    \caption{Distribution of the initial parameters. The masses of the stellar binary are defined as $m_1$ and $m_2$ ($m_1$ is more massive) with orbital elements $a_1$ and $e_1$. The stellar binaries orbit around the SMBH is defined with the orbital elements $a_2$ and $e_2$. 
    }
    \label{fig:ICs}
\end{figure*}

In \citet{Hoang+22}, Monte Carlo simulations of massive stellar binaries within $0.1$~pc of the GC's SMBH were implemented to explore the effects of natal kicks on the binaries. In this work, we expand on these earlier simulations and explore the effects that varying the initial stellar distribution has on the overall compact object density profile within the central parsec of the GC. See Figure \ref{fig:Possible_Configs} for a schematic of the methodology.

\subsection{Birth Configurations}
Each system begins as a hierarchical triple, comprising an inner binary of two main sequence stars ($m_1$ and $m_2$) and an outer binary consisting of the orbit around an SMBH. The frame of reference is selected to be the invariable plane and we define the orbital parameters of the inner (outer) binary using the Keplerian elements for the semimajor axis, $a_1$ ($a_2$), eccentricity, $e_1$ ($e_2$), inclination, $i_1$ ($i_2$), argument of periapsis, $\omega_1$ ($\omega_2$), longitude of the ascending node, $\Omega_1$ ($\Omega_2$), and true anomaly, $f_1$ ($f_2$). The inner and outer orbits are inclined to each other by a mutual inclination, $i_{tot} = i_1 + i_2$.

We define $m_1$ to be the more massive stellar binary member, such that it is always the first to undergo a supernova (SN) explosion. The mass distribution of $m_1$ is chosen from a Kroupa IMF ranging from 8 - 100 ${M}_\odot$ \citep{Kroupa01}. The mass ratio, defined as $q = m_2/m_1$, is chosen from a uniform distribution ranging from 0.1 - 1 \citep{Sana+12}. We set the mass of the SMBH at $m_{\bullet} = 4 \times 10^6 \: {M}_\odot$  \citep[e.g.,][]{Ghez+05, Gillessen+09}.

The eccentricity distribution for the inner binary $e_1$ is uniformly distributed between 0 and 1, while the outer orbit eccentricity $e_2$ is taken from a thermal distribution \citep{Jean+19}. The mutual inclination $i_{tot}$ between the inner and outer orbit is distributed isotropically. The argument of periapsis, true anomalies, and the inner binary longitude of ascending node are selected from a uniform distribution between $0$ and $2 \pi$. 

We choose the outer semimajor axis $a_2$ to follow a power-law density cusp, $n \propto r^{-\alpha}$, with a minimum semimajor axis of $500$~au and a maximum of $1$~pc. We vary $\alpha$ across the range of 0 to 3, in half integer increments and for each value of $\alpha$, we run $1.5$ million Monte Carlo simulations of the stellar binary orbiting around the SMBH.

The semimajor axis of the inner binary $a_1$ is determined from the period distribution $dn/dP \propto log(P)^{-0.45}$ \citep[]{Sana+13}{}, with the minimum and maximum value for $a_1$ selected for each system according to the following conditions:

\begin{itemize}

    \item First, we require that the stellar binaries' orbit pericenter be greater than two times the Roche limit of the system to ensure the stellar binary is not disrupted prior to the first natal kick: 
    \begin{equation}
        a_{1}(1-e_1) > 2a_{Roche} .
        \label{eq:BinRoche}
    \end{equation}
    The Roche limit of the stellar binary defined as 
    \begin{equation}
        a_{Roche, ij} = \frac{R_j}{\mu_{Roche, ji}} ,
    \end{equation}
    where $R_j$ is the radius of the star at mass $m_j$ and $\mu_{Roche, ji}$ is the approximation of the Roche lobe radius \citep[]{Eggleton83}{}{}:
    \begin{equation}
        \mu_{Roche, ji} = \frac{0.49 \, (m_j/m_i)^{2/3}}{0.6\, (m_j/m_i)^{2/3} + \ln(1 + (m_j/m_i)^{1/3})}.
    \end{equation}

    \item The upper limit for the $a_1$ distribution comes from ensuring that the system is hierarchically stable \citep[][]{Naoz16}:
    \begin{equation}
        \frac{a_1}{a_2} \frac{e_2}{1-e^2_2} < 0.1  .
    \label{eq:TripleStability}
    \end{equation}

    \item Finally, each triple system must also satisfy the following criteria of the stellar binary system not crossing the Roche limit of the SMBH before $m_1$ undergoes a supernovae explosion:
    \begin{equation}
        a_2(1-e_2) > a_1(1+e_1)\Bigl(\frac{3m_{\bullet}}{m_1 + m_2}\Bigl)^{1/3} .
    \label{eq:SMBHDisruption}
    \end{equation}

\end{itemize}

\subsection{Binary Destruction}
\label{Binary Destruction}
The initial stellar binaries can be destroyed either before or after the supernova. We track merged and unbounded stellar binary members in our simulation. Therefore, our simulations consist of a population of binary and single-star systems orbiting the SMBH. There are three paths to destroying the binary before either star has gone supernova: 
\begin{itemize}
    \item {\it SMBH Roche limit crossing.} $32 - 46\%$, from $\alpha=0-3$, respectively, of the initial stellar binaries distribution (see Figure \ref{fig:ICs}) did not meet Equation (\ref{eq:SMBHDisruption}) criterion. These evolve independently as single stars orbiting the SMBH. In the statistical analysis below, we incorporate both the single star population and the binary star population.
    \item {\it Stellar mergers induced by EKL.} A fraction of stellar binaries will experience eccentricity oscillations induced by the eccentric Kozai-Lidov mechanism \citep[EKL;][]{Naoz16} and can become a merged stellar product before the first natal kick \citep[e.g.,][]{Antonini+14,Prodan+15,Stephan+16,Wang+21}. Following \citet{Hoang+22}, we incorporate a simplified condition for which systems that have an EKL timescale shorter than general relativity (GR) precession may merge (or at least undergo mass transfer). We find that roughly {$1 - 6\%$}, from $\alpha=0-3$, respectively, of the initial stellar binaries fall into this category and are excluded from undergoing supernovae explosions in our simulations. 
    \item {\it Unbinding via neighboring scattering interactions (evaporation).} Weak gravitational interactions with nearby stars can unbind the binary over an evaporation timescale \citep[e.g.,][]{BT}: 
    \begin{equation}
        t_{evap} = \frac{\sqrt{3} \: \sigma(r)}{32 \sqrt{\pi} G \rho(r) a_1 ln(\Lambda)} \frac{m_1 + m_2}{m_p} ,
    \end{equation}
    where {$ln(\Lambda) = 5$} is the Coulomb logarithm \citep{Rose+20}, $m_p$  is the average mass of the perturbing star, $\sigma(r)=\sqrt{Gm_{\bullet}/r(1+\alpha)}$  and $\rho(r)$, is defined below. Note that for simplicity, we ignore the eccentricity of the binary about the SMBH, since it will only change the timescale by a factor of a few \citep{Rose+20}. 

We point out that we are testing a wide range of density profiles, $\alpha=0-3$, see Equation (\ref{eq:rho}). However, observations of the galactic center suggest a shallow, core-like profile \citep[$\alpha \sim 1.1 - 1.4$][]{Schodel+18,Gallego-Can+18}. Thus, following \citet{Gallego-Can+18} and \citet{Rose+20}, we adopt the evaporating population distribution to be with $\alpha=1.3$.  In this case, most binaries have an evaporation timescale longer than the supernova timescale for a range of separations about the SMBH \citep[e.g.,][]{Rose+20}. Only about {$7 - 9\%$} of the remaining stellar binaries, from $\alpha=0-3$, respectively, will evaporate before the first supernova. \footnote{Assuming that the profile of all the stellar components in a nuclear star cluster follows the adopted density profile. In this case, {$10 - 24\%$} of the remaining stellar binaries from $\alpha=0-3$, respectively, will evaporate before the first supernova. }.

    
    
\end{itemize}

\begin{table*}[t]
\centering
\begin{tabular}{ccccccccc}
\toprule
{$\alpha$} & \multicolumn{3}{c}{Initial} & \multicolumn{5}{c}{Post-Kick}     \\
\cline{2-4}
\cline{6-8}
{} &  EKL (\%) &    Singles (\%) &  Binaries (\%) &   {}  &  {}  &  Singles (\%) &    Binaries (\%)  \\

\midrule
\midrule
\textbf{0.0        } &    {$1.0$} & {$40.8$} & {$58.2$} &  & & $94.5$  & $5.5$ & \\
\textbf{0.5        } &    {$1.0$} & {$41.0$} & {$58.0$} &  & & {$94.5$}  & {$5.5$} & \\
\textbf{1.0        } &    {$1.1$} & {$41.5$} & {$57.4$} &  & & {$94.6$}  & {$5.4$} & \\
\textbf{1.5        } &    {$1.3$} & {$42.1$} & {$56.6$} &  & & {$94.6$}  & {$5.4$} & \\
\textbf{2.0        } &    {$1.7$} & {$43.5$} & {$54.8$} &  & & {$94.7$}  & {$5.3$} &  \\
\textbf{2.5        } &    {$3.0$} & {$46.9$} & {$50.1$} &  & & {$94.9$}  & {$5.1$} & \\
\textbf{3.0        } &    {$5.9$} & {$54.1$} & {$40.0$} &  & & {$95.5$}  & {$4.5$} & \\
\bottomrule
\end{tabular}
\caption{Percentage of initial systems and post-kick systems in the binary or single configuration. Initial systems in the single configuration are the result of binaries being disrupted by one of the processes that occur before the first natal kick, described in Section \ref{Binary Destruction}.  }
\label{BinariesAndSingles}

\end{table*}

The remaining inner binaries can also be destroyed at a later time due to natal kicks or close encounters with the SMBH. Because $m_1$ is the more massive companion, it will undergo a supernova explosion first. The first natal kick can disrupt the binary, leading to the formation of two separate orbits around the supermassive black hole ($m_1$ - SMBH and $m_2$ - SMBH). If the binary survives $m_{1}$'s natal kick, then we are left with a binary consisting of a compact object (CO) and star orbiting the SMBH. This scenario may result in the formation of X-ray binaries (XRB, Section \ref{XRB}). $m_{2}$'s natal kick provides an additional way of destroying the binary and for the creation of Gravitational wave mergers (GW mergers, Section \ref{GWMerg}). Either natal kick can also push the binary onto an orbit inside the SMBH Roche limit, resulting in the destruction of the binary.

\subsection{Pre-Supernova Evolution}
Each star in the inner binary experiences mass loss due to main-sequence evolution. Between birth and the first supernova, the inner and outer binary will expand due to mass loss but the outer binary will expand by a negligible amount because of the large mass of the SMBH. Using the rapid single stellar evolution code SSE \citep{Hurley+00}, we determine the time that each star becomes a CO and the  mass prior to and following this event. By adopting adiabatic expansion, which conserves angular momentum, the inner binary semimajor axis immediately before the first supernovae, {$a_{1,pre-SN}$} is: 
\begin{equation}
    a_{1,pre-SN} = \frac{m_1 + m_2}{m_{1, pre-SN} + m_{2, pre-SN}} a_1
    \label{eq:massloss}
\end{equation}
where $m_{1, pre-SN}$ and $m_{2, pre-SN}$ are the masses of $m_1$ and $m_2$ immediately before the first supernovae.

\subsection{Applying Supernovae Kicks}
We assume instantaneous supernovae kicks that are isotropically distributed. Supernovae kicks for NSs are selected from a normal distribution with an average of $400$~km~sec$^{-1}$ and standard deviation of $265$~km~sec$^{-1}$ \citep[e.g.,][]{Hansen+Phinney97, Arzoumanian+02, Hobbs+04}. We adopt two different BH kick prescriptions due to observational uncertainties. In the fast BH kick prescription, BHs have the same kick distribution as NSs. In the slow BH kick prescription, the BHs receive the same linear momentum kick as NSs \citep{Bortolas+19}. 

Recent studies suggest the possibility that NSs receive smaller birth kicks if they are formed through the electron-capture supernova (ECSN), for a mass range of 6-10$M_\odot$ \citep{Miyaji+1980, Nomoto+1987, Poelarends+2008, Jones+2016, Leung+2020}. However, only a window of approximately $\sim 0.2 M_\odot$ in the 6-10 $M_\odot$ mass range actually undergoes electron capture kicks \citep{Doherty+2017, Willcox+2021, Hiramatsu+2021, Stevenson+22}. We also conducted two separate numerical experiments, considering ECSN. One for which all the stars between 6-6.2 $M_\odot$ underwent ECSN, or $\sim 5\%$ of the entire population. Thus, the inclusion of ECSN at face value seems negligible. On the other hand, an extreme case is the one in which all NS progenitors undergo ECSN. We chose an ECSN kick distribution that will lead to the maximum variation by taking a Maxwellian with $\sigma = 30$~km/s. The numerical experiment, in this case, is consistent with the slow-BH kicks, where the post-kick density profiles have nearly identical slopes and similar amounts of unbound systems. This result is insensitive to the particular choice of ECSN kick distribution \citep[e.g.,][]{Willcox+2021, Gessner+2018, Stevenson+22}. We, thus, omit the results from these experiments to avoid clutter throughout the paper.

$m_1$ will undergo a supernova explosion first because it is the more massive companion. The supernova kick is applied by adding the Cartesian velocity kick vector to the orbital velocity vector of $m_1$ and changing $m_1$ to the post-supernova mass found with {\tt SSE}. Following the first supernova, there are two main scenarios: The inner binary survives $m_1$'s supernova kick or is disrupted by $m_1$'s supernova kick.

In the scenario where the inner binary survives, it can remain bound to the SMBH on an elliptical orbit or become unbound from the SMBH on a hyperbolic orbit. Just before $m_2's$ supernova kick, we adiabatically expand the orbits due to mass loss from $m_2$ using Equation \ref{eq:massloss}.

For elliptical orbits, if the timescale between the first and second supernova kick exceeds 10 times the orbital period, we randomly select the eccentric anomaly at the time of $m_2's$ kick from a uniform distribution between $0$ and $2 \pi$. Otherwise, we determine the eccentric anomaly by iteratively solving the elliptical Kepler's equation using Newton's method. For hyperbolic orbits, we solve the hyperbolic Kepler Equation using the HKE-SDG package \citep{Rapose2018} to find the hyperbolic anomaly at the time of $m_2$'s kick. With either the eccentric or hyperbolic anomaly, we calculate the true anomaly, and then determine the Cartesian coordinates of the orbit immediately prior to $m_2's$ kick.

With the calculated Cartesian coordinates of the orbit, $m_2$'s supernova kick is applied by adding the Cartesian velocity kick vector to the orbital velocity vector of $m_2$ and changing $m_2$ to the post-supernova mass found with {\tt SSE}.

In the second scenario that the inner binary is disrupted by the first supernova kick, $m_1$ and $m_2$ form separate binaries with the SMBH. Then $m_2$'s supernova kick is applied by adding the Cartesian velocity kick vector to the orbital velocity vector of $m_2$ and changing $m_2$ to the post-supernova mass found with {\tt SSE}.

\subsection{Interaction with the SMBH}

If the separation of the inner binary (either progenitor or post-kick binary) is larger than the SMBH's Roche limit, Eq.~(\ref{eq:SMBHDisruption}), then the binary is disrupted and we follow the individual star's evolution. Further, binaries disrupted by natal kicks form two separate orbits around the supermassive black hole ($m_1$ - SMBH and $m_2$ - SMBH). If the binary is disrupted by $m_{1}$'s natal kick, then it is possible that $m_{2}$ will be on an orbit that will create a tidal disruption event (TDE, Section \ref{TDE}). On the other hand, if the binary is disrupted after the second kick, the result may lead to an extreme mass ratio inspiral (EMRI, Section \ref{EMRIS}).

\subsection{Normalization}
Throughout this paper, we normalize the density distribution by the $M-\sigma$ relation \citep{Tremaine+02}:
\begin{equation}\label{eq:rho}
    \rho(r) = \frac{3-\alpha}{2\pi} \frac{m_1}{r^3}\left(\frac{G\sqrt{m_1 M_0}}{\sigma_0^2 r}\right)^{-3+\alpha} ,
\end{equation}
where $M_0=10^8$~M$_\odot$, and $\sigma_0=200$~km~sec$^{-1}$. In the rest of this paper, we refer to the numbers of NSs and BHs as expected from this normalization process. Here, we can also recognize a notable quantity called the ``sphere of influence,'' which signifies the radius at which the gravitational potential is dominated by the SMBH's. Equation (\ref{eq:rho}) implies that this value is: $r_h=G\sqrt{m_1M_0}/\sigma_0^2$, in our own GC.

\section{Dark Cusp and High Energetic Phenomena Predictions }
\label{Sec:PredictionsATGC}

\subsection{The Relationship between Dark Cusp and stellar density distribution}
\label{Sec:DarKCusp}

\begin{figure}
    \centering
    \includegraphics[width=9cm]{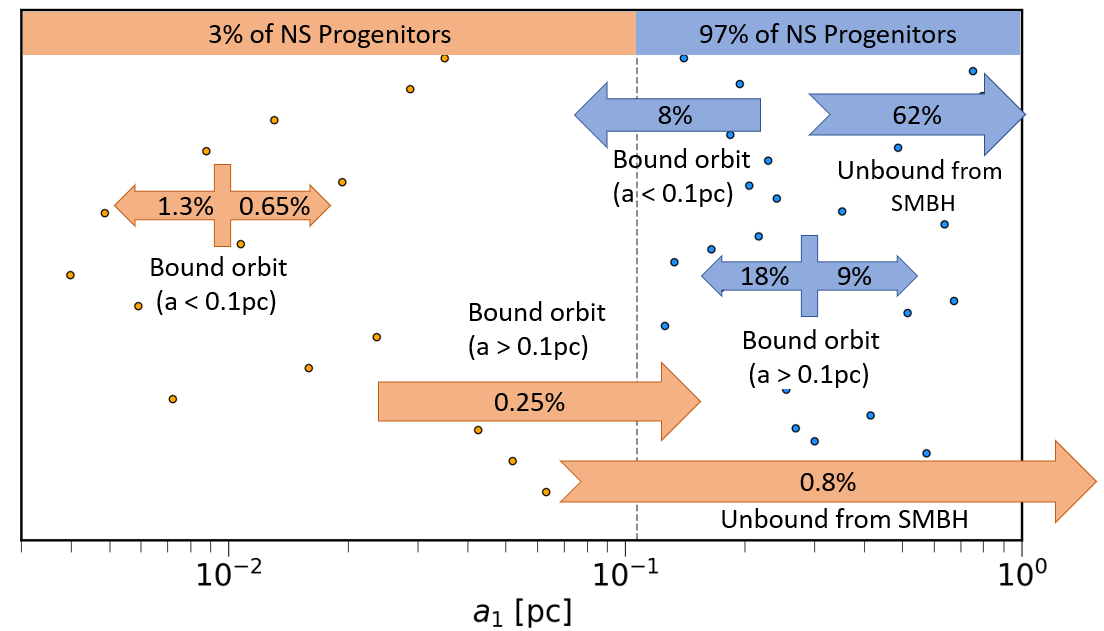}
    \caption{Schematic representation of the possible movement of NSs due to the natal kicks with $\alpha$ set to 1.5. At 0.107pc, the average kick velocity ($\simeq$ 400 km/s) is equal to the circular orbital velocity and this location is denoted by the dashed grey vertical line. The orange (blue) dots represent the NS progenitors located within 0.1pc (between 0.1pc - 1pc). $3\%$ ($97\%$) of all the NS progenitors are formed within 0.1pc (0.1 - 1pc).}
    \label{fig:schematic}
\end{figure}

\begin{figure*}[t]
    \centering
    \includegraphics[width=18cm]{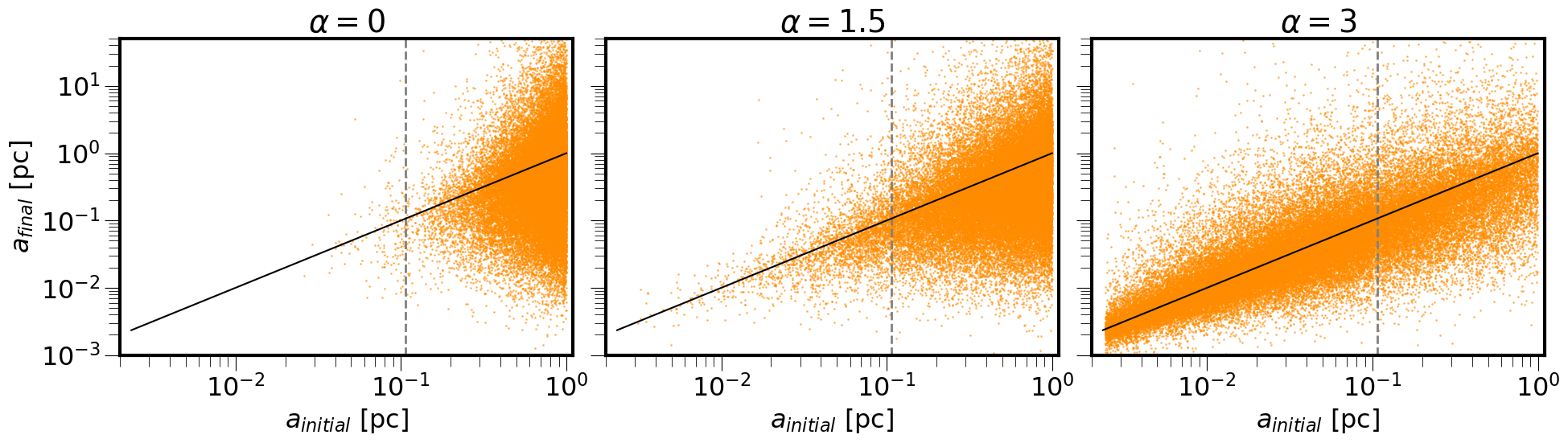}
    \caption{{\bf Three examples of the steepening of the neutron star density.} Here we show the final vs. the initial semimajor axis of the NSs for three representative initial density distributions. Specifically, we consider a shallow ($\alpha=0$, left panel) and steep ($\alpha=3$, right panel) distributions. We also present an intermediate distribution of $\alpha=1.5$ \citep[close to the observed stellar distribution, middle panel][]{Gallego-Can+18}. The black line represents the line that NSs would remain on provided that there were no natal kicks. As depicted, in the $\alpha=0$ and $\alpha=1.5$, the initial inner parts of the parameter space are almost entirely devoid of NSs progenitors. Post kick, about $\simeq 6$~\% ($\simeq 8$~\%) of the total progenitors' population of the $\alpha=0$ ($\alpha=1.5$), moved inwards of $0.1$~pc.  For the $\alpha=3$ case, where the distribution is initially constant in $\log a_{\rm initial}$, not much changed, post kick. This behavior is further illustrated in Figure \ref{fig:NS Density}.}
    \label{fig:Positon}
\end{figure*}

\begin{figure*}[t]
    \centering
    \includegraphics[width=18cm]{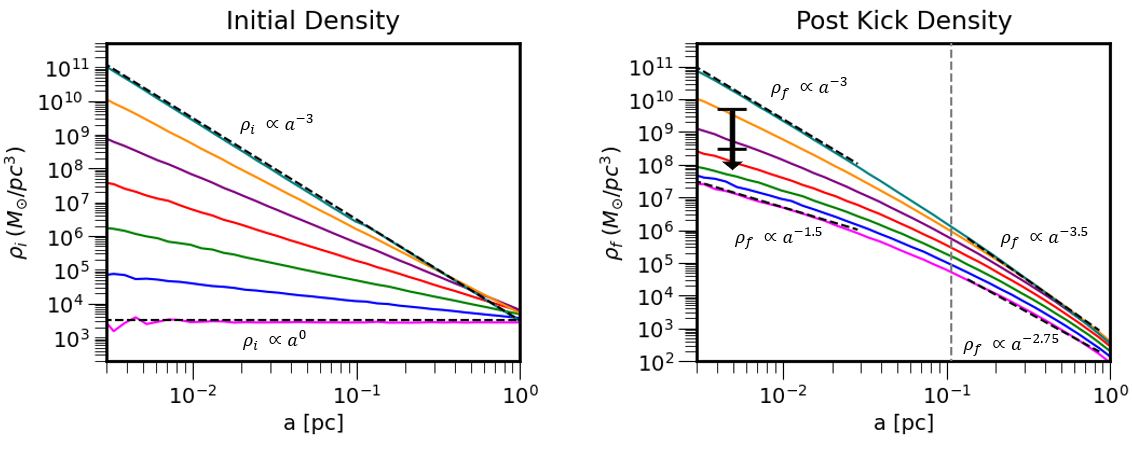}
    \caption{Density profile of NS progenitors (left panel) and the bound NS (right panel) after the natal kicks as a function of semimajor axis.  The post-kick density power law slope has a break at the characteristic location of $a = 0.107$~pc, where the circular orbital velocity equals the average supernovae kick speed. We consider from bottom to top the following density profiles $\alpha= 0 - 3$. Note that the density profiles, become steeper post-kick (see text for details). The uppermost horizontal black line at $a = 4 \cdot 10^{-3}$ pc indicates the upper limit of the enclosed mass within S0-2's orbit with all of the mass assumed to be in NSs \citep[e.g.,][]{Heibel2022, Gravity+22}. The lower black line is the upper limit assuming a typical NS population fraction of $0.26:0.014:2.3\times 10^{-3}$ of WD:NS:BH \citep{Kroupa01}. 
    }
    \label{fig:NS Density}
\end{figure*}

The various dynamical processes described in the Section above disrupt a significant fraction of binaries before the first supernova. The natal kicks disrupt the majority of the remaining binaries, and by the end of the simulations, only a small fraction of all initial binaries remain bound to their companion (see Table \ref{BinariesAndSingles} for details). The majority of the systems are single COs that are either orbiting the SMBH or unbound from the SMBH. In general, the COs do not remain in their initial position and are scattered.

There are two significant outcomes for a single or binary configuration post-kick. One is if the binary or single remains bound to the SMBH, meaning the configuration post-kick has Keplerian energy smaller than zero. The other is to become unbound to the SMBH; in other words, the Keplerian energy is larger than zero. Out of these $\simeq 20\%$ are on a trajectory to escape the galaxy. 

A schematic description of this result is depicted in Figure \ref{fig:schematic}, where we show an example for $\alpha=1.5$, which is a core-like distribution similar to the one observed in our GC \citep[e.g.,][]{Schodel+18,Gallego-Can+18}. Although only $3\%$ of the NS progenitor population is formed within $0.1$~pc, natal kicks move NSs that were originally located at a distance $>0.1$~pc toward the GC, and ultimately $9.3\%$ of NSs end up within $0.1$~pc. 
At $r = 0.107$pc, the average kick velocity ($\simeq 400$~km/s) is equal to the circular orbital velocity and serves as a critical point for differentiating the behavior of the NS population in the two regions. $26.7\%$ ($64\%$) of the NSs initially formed with a semimajor axis less (greater) than $0.1$~pc are unbound from the SMBH. The combination of this, along with a steepening of the NS number density within the $0.1$~pc threshold, leads to a dense concentration of NSs within the $0.1$~pc radius and a scarcity beyond it. 

Below we highlight a few observational tests that can be used to constrain the CO progenitors' stellar distribution due to the unique nature of the GC and the natal kicks. Natal kicks efficiently move COs closer to the SMBH. Thus, observational constraints of the dark cusp may be used to constrain the initial stellar distribution. 




Future observations can be used to constrain the dark cusp. The separation of a young binary at the inner 0.1 pc of the GC is sensitive to the underlying density profile and measurements of such systems could be used to place constraints on the dark cusp \citep[e.g.,][]{Alexander05,Rose+20}.

\subsection{The effect of progenitor distribution on the post-kick density and eccentricity distribution }

\begin{figure*}
    \centering
    \includegraphics[width=18cm]{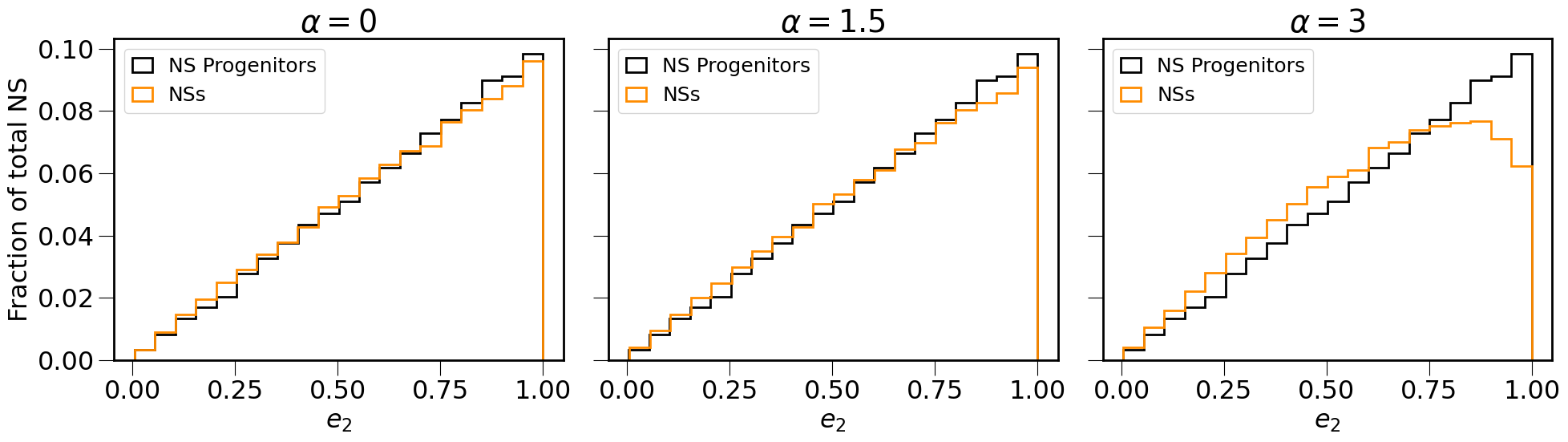}
    \caption{ \textbf{Three examples of the eccentricity changes of the bound neutron star population.} Following Figure \ref{fig:Positon}, we include a shallow, intermediate, and steep initial stellar distribution for $\alpha=0,1.5,3$ respectively. In the shallow and intermediate cases, natal kicks results in a slight decreased population of highly eccentric orbits. }
    \label{fig:NS_Eccentricity}
\end{figure*}

Below we provide a detailed analysis of the NS distribution. The fast kick BH distribution follows the NS distribution (only with a different normalization). The slow BH kick results are described in Appendix \ref{BH Distribution}. 

In Figure \ref{fig:Positon}, we show the changes in the bound NSs semimajor axis due to the kicks for three different density profiles, from extremely shallow ($\alpha=0$, left), extremely steep ($\alpha=3$, right), as well as core-like distribution closer to the observed distribution ($\alpha=1.5$, middle). As depicted, NS progenitors formed near $1$~pc can move orders of magnitudes away from their birth positions while those formed in the nearby vicinity of the SMBH are scattered by only an order of magnitude or so. The shallowest initial density profiles (i.e., $\alpha = 0, 0.5$) contain the majority ($\simeq 99\%$) of the NS population outside of $0.1$~pc and so are significantly perturbed by the NS kicks and steepen dramatically within $a = 0.107$~pc. As the value of $\alpha$ increases, a larger fraction of NSs are initially within $0.1$~pc of the SMBH, and so the increase in steepness is less susceptible to natal kicks, as further demonstrated in Figure \ref{fig:NS Density}.

In Figure \ref{fig:NS Density}, we show the NS progenitor (left panel) and bound NS (right panel) density distributions after the natal kicks. The bound NS density profiles are all steeper than their corresponding progenitor profiles. As the initial progenitor profiles increase in steepness, the corresponding amount of steepening in the bound profile decreases. The post-kick density profile can be estimated analytically from the number of systems that become unbound to the SMBH. Conservation of particles implies that the main driver of the post-kick distribution is the fraction of systems remaining. We provide the details in Appendix \ref{CO Density Slopes Calc}

We apply a density criteria on the NS density profiles to constrain the expected initial stellar profile from observations of the precession of S0-2s caused by the unseen mass within S0-2s orbit \citep[e.g.,][]{Do+19, Gravity+22,Heibel2022}. The upper limit is derived by assuming that all of the enclosed mass is NSs. In this case, an initial stellar profile with $\alpha < 3$ is consistent with this constraint. However, if there are also white dwarfs and stellar-mass black holes in this vicinity, assuming the typical population fraction of $0.26:0.014:2.3\times 10^{-3}$ of WD:NS:BH \citep{Kroupa01}, means that about $\simeq 5\%$ of the unseen mass is in NSs. Then, an initial stellar profile of $\alpha > 2$ is incompatible with mass constraints. Further observational measurements may be able to disentangle the mass fraction of NSs within S0-2's orbit and provide a more stringent test on the initial stellar profile.   




Lastly, the kicks may also significantly affect the NS eccentricity distribution, especially for extremely cuspy density profiles. Initially, all CO progenitors begin on a thermal eccentricity distribution. Note that a thermal distribution may not accurately describe the eccentricity distribution at the GC \citep{Geller+19} but is used here as a proxy.  In Figure 5, we display the changes in NS eccentricity due to the kicks for three different density profiles with a shallow ($\alpha = 0$), intermediate ($\alpha = 1.5$), and steep ($\alpha = 3$) distribution. For shallow initial stellar distributions ($\alpha = 0, 1.5)$, the post-kick eccentricity distribution follows the initial thermal distribution at lower eccentricities and drops slightly when $e > 0.7$. 
When considering the steeper distribution near $\alpha = 3$, the orbits tend toward circularization, resulting in a higher proportion of orbits characterized by low eccentricities.

\subsection{Extreme Mass Ration Inspirals (EMRIs)}
\label{EMRIS}
Extreme Mass Ratio Inspirals (EMRIs) are GW emission events that take place when stellar mass COs inspirals onto SMBHs. They are one of the prime science motivators of the future Laser Interferometer Space Antenna (LISA) and other mHz detectors \citep[e.g.,][]{Amaro+17}. Natal kick can drive a CO into the SMBH \citep[e.g.,][]{Bortolas+17, Hoang+22}. To estimate if a kick resulted in an EMRI we compare two timescales. One describes the characteristic GW decay timescale
\begin{equation}
    t_{\rm GW, EMRI} \simeq \frac{5}{64} \frac{c^5a^4}{G^3m^2_{\bullet}m}(1-e^2)^{7/2} ,
\end{equation}
where $c$ is the speed of light, $e$ is the eccentricity of the object around the SMBH, post-kick, and $a$ is its semimajor axis \citep[][]{Peters64}. The other timescale is two-body relaxation $t_{\rm relx}$ which is the result of weak kicks with other neighboring objects. On one hand, these kicks can result in EMRIs by changing the angular momentum of the orbit and driving it into the lost cone. On the other hand, the kicks can increase the angular momentum, yielding a more circular orbit and thus suppressing the formation of an EMRI. Following \citet{Amaro+07}, we classify an orbit to be an EMRI if $t_{\rm GW, EMRI}<(1-e) t_{\rm relx}$ is satisfied. 

\begin{figure}
    \centering
    \includegraphics[width=8cm]{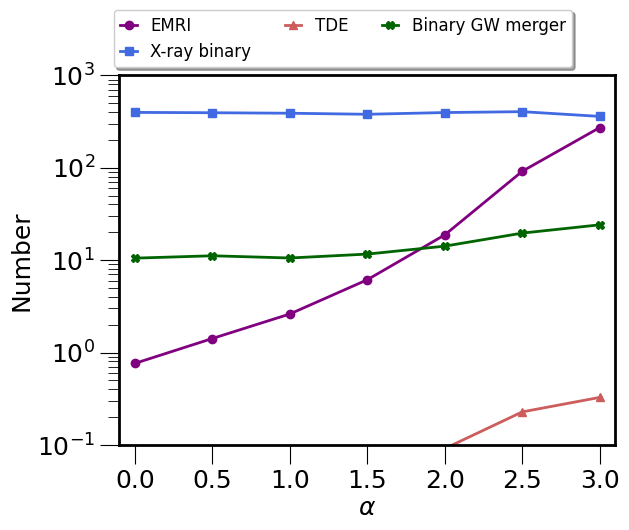}
    \caption{Number of transient observables within the sphere of influence of the SMBH. We classify the observables in the following sections: EMRIs (Section \ref{EMRIS}, combining BH and NS EMRIs together), X-ray binaries (Section \ref{XRB}), TDEs (Section \ref{TDE}), Binary GW mergers (Section \ref{GWMerg}). The number here represents the expected number adopting the $M-\sigma$ relation after one star formation episode.  }
    \label{fig:Observables_at_SOI}
\end{figure}

We convert the number of EMRIs in our simulations to the number of EMRIs within the sphere of influence, as expected from the $M-\sigma$ relation. As shown in Figure \ref{fig:Observables_at_SOI}, we find that EMRI formation is sensitive to the initial stellar distribution surrounding the SMBH. In particular, the expected number of EMRIs range from nearly 0 EMRIs for a shallow cusp ($\alpha = 0$) to $270$ EMRIs for a steep cusp ($\alpha = 3$). Considering a stellar profile that closely resembles the one observed in the GC \citep{Gallego-Can+18}, we expect less than 10 EMRIs driven by natal kicks. For all initial stellar profiles, the majority of EMRI progenitors are formed within $10^{-1}$~pc and are the result of NSs inspiraling onto the SMBH. We find that $98\%$ $(92\%)$ are NS EMRIs and $2\%$ $(8\%)$ are BH-EMRIs, for alpha = 0 (3). We note that for $\alpha\leq2$ the expected number of EMRIs from this channel is lower than the expected number of EMRIs from two body relaxation \citep[e.g.,][]{Hopman+Alexander06,Sari+Fragione19}, and orders of magnitude lower than the expected number of EMRIs in SMBH binaries \citep{Naoz+22,Naoz+23}. For the extreme cusp case, i.e., $\alpha\geq 2.5$, the expected numbers combined NS and BH EMRIs are comparable to the lower limit of the SMBH binary case. We suggest that extreme cusp profiles may also contribute to the
 revised stochastic background estimations presented in \citet{Naoz+23}. We reserve this calculation for future studies.


\subsection{X-Ray Binaries}
\label{XRB}
Inner binaries that survive $m_1$'s natal kick can have their orbital separation decrease. Following \citet{Naoz+16}, we classify systems as X-ray binaries if the inner binary post-kick pericenter drops below $a_{Roche}$. We note that a binary system crossing the Roche limit is a necessary, yet insufficient condition for its evolution into an X-ray binary. The transformation into an X-ray binary also depends greatly on the specific evolutionary characteristics of the secondary star. Therefore, we can provide an upper limit on the number of X-ray binaries created from natal kicks.

We find that $3.3 \cdot 10^{-3}$ NS X-ray binaries form per NS and $1.2 \cdot 10^{-3}$ BH X-ray binaries form per BH in our simulations for all values of $\alpha$ other than $\alpha = 3$. There is a decrease in the X-ray binary fraction for $\alpha = 3$ because there is a significant decrease in the number of initial stellar binaries (see Table \ref{BinariesAndSingles}). We find that the formation of X-ray binaries are related to the properties of the inner binary and is independent of the binary's outer orbital parameters, such as distance away from the SMBH. From Figure \ref{fig:Observables_at_SOI}, we expect nearly 400 to be formed within the sphere of influence due to natal kicks. From this, 94\% of the X-ray binaries are NS X-ray binaries, and 6\% are BH X-ray binaries. The high abundance of X-ray sources observed at the galactic center \citep[e.g.,][]{Hailey+18,Zhu+18} might be explained by these X-ray binaries.

\begin{table*}
\centering
\begin{tabular}{ccccccc}
\toprule
{$\alpha$} & \multicolumn{2}{c}{NSs} & \multicolumn{4}{c}{Slow Kick BHs}     \\
\cline{2-3}
\cline{5-7}
{} &      SMBH-bound (\%) &    SMBH-unbound (\%) &&&    SMBH-bound (\%) &    SMBH-unbound (\%)  \\
\midrule
\midrule
\textbf{0.0        } &    32.1 &  67.9 &&&    83.1 &  16.9 \\
\textbf{0.5        } &    33.2 &  66.8 &&&    83.7 &  16.3 \\
\textbf{1.0        } &    34.8 &  65.2 &&&    84.6 &  15.4 \\
\textbf{1.5        } &    37.4 &  62.6 &&&    85.6 &  14.4 \\
\textbf{2.0        } &    42.5 &  57.5 &&&    87.8 &  12.2 \\
\textbf{2.5        } &    53.4 &  46.6 &&&    91.2 &  8.8  \\
\textbf{3.0        } &    70.5 &  29.5 &&&    95.7 &  4.3  \\
\bottomrule
\end{tabular}
\caption{Percentage of COs who are bound (unbound) from the SMBH after the natal kicks for values of $\alpha$.}
\label{CO_Bound_Unbound}
\end{table*}

\subsection{Tidal Disruption Events (TDEs)}
\label{TDE}
Tidal disruption events (TDEs) occur when $m_1$'s natal kick disrupts the stellar binary and pericenter of the $m_2 - SMBH$ orbit drops below the SMBH tidal radius 
\begin{equation}
    r_t \sim R_{*}\left( \frac{m_\bullet}{m_{*}} \right)^{1/3} ,
\end{equation}
where $R_{*}$ is the radius of the star and $m_{*}$ is its mass. We further require that $r_t$ is greater than the SMBH Schwarzschild radius and that $m_2$ passes within the tidal radius before its own natal kick to classify the system as a TDE. TDEs are a rare outcome of natal kicks acting on binaries. We find that no TDEs driven by natal kicks are expected to occur within the sphere of influence of the SMBH. 

TDEs are expected to result via two-body relaxation processes \citep[e.g.,][]{Rees88,Hopman+Alexander05,Fragione+Sari18,Madigan+18,Akiba+21}, an in SMBH binaries \citep[e.g.,][]{Chen+09,Chen+11,Mockler+23,Melchor+23}.



\subsection{Inner Binary GW mergers}
\label{GWMerg}
The natal kicks can also direct the surviving inner binaries into regions of the parameter space where GR effects trigger a gravitational wave (GW) merger within a timescale shorter than the evaporation timescale at the GC. The inner binary gravitational wave merger timescale due to GR effects is \citep{Peters+63}:
\begin{equation}
    t_{GW} \sim \frac{5}{265} \frac{c^5 a_1^4}{G^3 (m_1 + m_2)m_1m_2}(1-e^2)^{7/2} .
\end{equation}
We label a system as a GW merger if $t_{GW} < t_{evap}$. 
In some cases, the EKL-induced eccentricity oscillations play a significant part in inducing a GW merger. If the EKL timescale is shorter than the GR precession timescale, we describe the EKL-induced GW merger timescale as 
\begin{equation}
    t_{GW_{EKL}} \sim \frac{5}{265} \frac{c^5 a_1^4}{G^3 (m_1 + m_2)m_1m_2}(1-e^2_{1,max})^{3} ,
\end{equation}
where $e_{1,max}$ is the maximum EKL-induced eccentricity and is estimated following \citet{Wen}. GW mergers are weakly dependent on the assumed initial stellar distribution and will result in $10 - 25$ GW mergers within the sphere of influence of the SMBH.

\section{Predictions for the Roman Space Telescope}
\label{Sec:PredictionForRoman}



\subsection{Compact Object Distribution beyond 1~pc}


\begin{figure*}
    \centering
    \includegraphics[width=16cm]{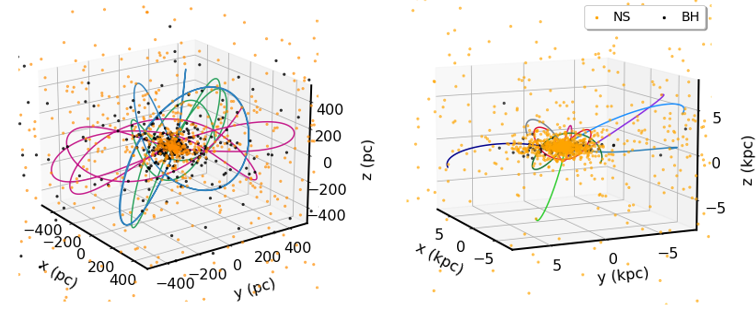}
    \caption{3D distribution of NS and BH population, with $\alpha=1.5$, after $100$~Myr from the initial star formation episode. The light dots mark a sample of NSs while the black dots are all of the BHs with slow kicks that ejected from the central SMBH. The colored lines are the orbits for a few selected COs. Note that the fast kick BHs follow the NS distribution and are omitted from the plot to avoid clutter. The left (right) panel represents the population of stellar remnants within a radial distance of $500$~pc ($5$~kpc) from the GC. To avoid overcrowding, only 1 out of every 13 NSs are shown.}
    \label{fig:3DPopulation}
\end{figure*}

\begin{figure*}[t]
    \centering
    \includegraphics[width=16cm]{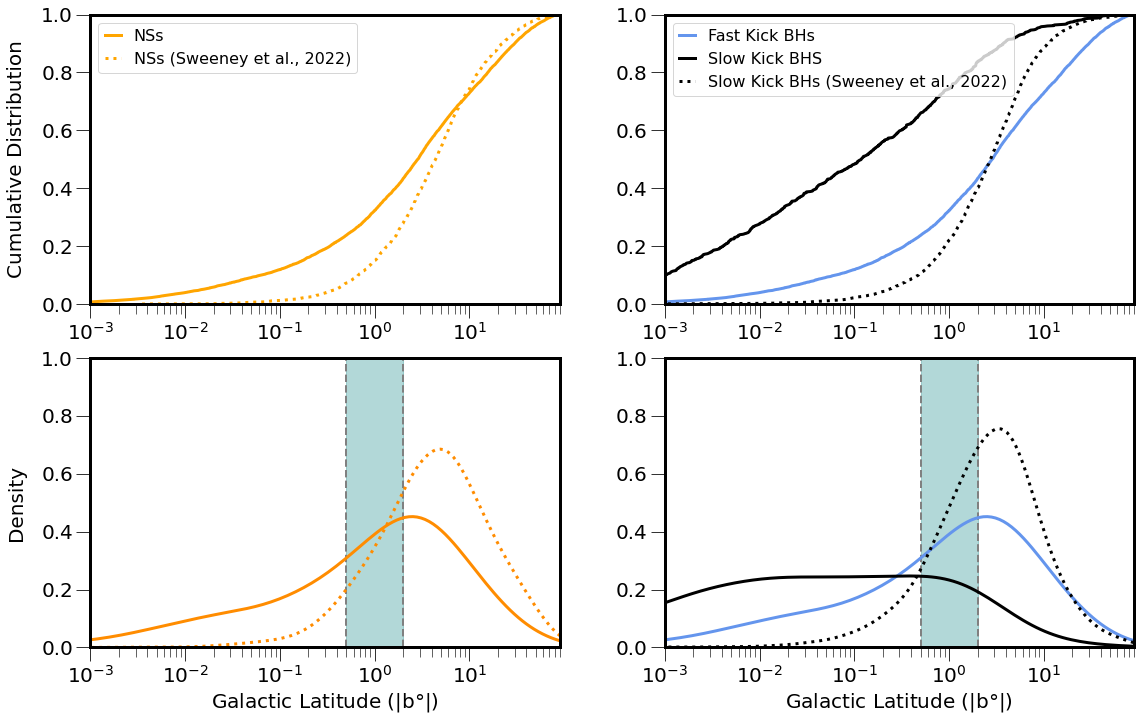}
    \caption{Galactic latitude distribution of COs ejected from the central parsec of the GC with an initial stellar profile of $\alpha$ = 1.5 (solid lines) after 3 Gyr. Also displayed are the distribution of post-kick COs from the entire galactic population (dashed lines), as analyzed by \citet{Sweeney+22}. In the top row, we plot the cumulative distribution function (CDF) for the NSs and BHs with the different kick prescriptions. In the bottom row, we plot the probability density functions for the COs. In both CO population sets, we limit the sample of COs in the distribution to be within a cylindrical radius of 8kpc from the GC. The teal shaded region shows the range of Galactic latitudes expected to be observed by the Roman Space Telescope \citep[e.g.,][]{Penny:2019}. }
    \label{fig:Roman_Distribution}
\end{figure*}

Consider a $3$~Gyr star formation episode within 1 pc of the SMBH \citep[consistent with][]{Zhou+23} \footnote{Note that a young stellar population at the GC is estimated to have an age of few Myrs \citep[e.g.,][]{Lu+09}, and while this population is interesting for its own merit, it provides negligible predicting power to the Roman Space Telescope. }. 
Within $1$~pc, all NS and BH progenitors are initially orbiting the SMBH, but the natal kicks unbind a significant fraction of COs from the SMBH potential, as described above (see Table \ref{CO_Bound_Unbound}). As expected, the percentage of COs that remain bound to the SMBH increases for a steeper initial stellar distribution. 

As a test case, we focus on the $\alpha = 1.5$ distribution. This density distribution is close to the GC observed stellar distribution \citep[e.g.,][]{Gallego-Can+18}, and agrees with the constraints in Figure \ref{fig:NS Density}. With $\alpha = 1.5$, $91\%$ of unbound systems are single COs (average speed of $\simeq 575$ km~s$^{-1}$) and $2\%$ are CO binaries (average speed of $\simeq 300$ km~s$^{-1}$). The remaining $7 \%$ are ejected during their stellar lifetime due to their companion's natal kick and will undergo their own supernova explosion outside the sphere of influence. These hyper velocity stars (average speed of $\simeq 600$ km~s$^{-1}$) can briefly be observed for $10^6 - 10^7$ years before becoming COs and contributing to the CO distributions. The combined gravitational potential of the Milky Way (MW) will be significant enough to slow down the majority ($\sim 70\%$) of systems unbound from the SMBH but bound to the MW potential with orbits scattered around the Galactic plane. Here we focus on those COs that remain bound to the MW after $3$~Gyr, and their potential detection using the Roman Space Telescope (Section \ref{RomanMicroLensing}).

We utilize the publicly available Python package for galactic dynamics {\tt galpy} \citep{Bovy2015}, to model a simple Milky Way potential. We follow the orbits of all (bound and unbound to the MW) COs beyond the inner $\sim 1$~pc. In Figure \ref{fig:3DPopulation}, we present the 3D distribution of a sample of COs ejected from the central parsec of the GC in a galactocentric coordinate frame. We display the position of COs and a few selected orbits $100$~Myr after the star formation event and within a radial distance of $500$~pc and $5$~kpc, respectively. In both panels, the orbits cross within the inner regions of the GC, consistent with what is expected for being expelled from this region and falling back into the MW potential. The slow BH kick prescription results in the BHs being concentrated closer to the GC than the NSs. The fast BH kick prescription results in the same density distribution of BHs and NSs, since by definition the fast BH kick prescription is matched to the observationally determined NS kick distribution.

\subsection{The Relation between Galactic Latitude and Kick Prescription}

\begin{figure*}[t]
    \centering
    \includegraphics[width=18cm]{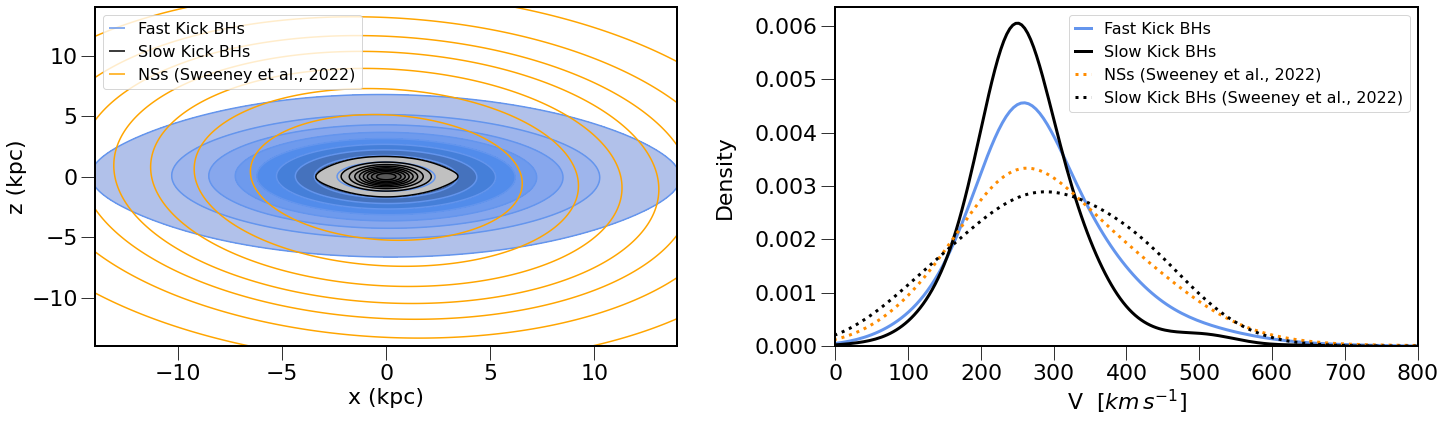}
    \caption{Spatial and velocity distribution of the BHs ejected from the central parsec and are bound to the Galaxy after 3 Gyr. In the left panel, we show the spatial distribution of the two BH populations in the galactocentric frame. In the right panel, we show the heliocentric speeds of the BHs for both BH populations. Similar to Figure \ref{fig:Roman_Distribution}, we compare the spatial and velocity distribution to the results obtained by \citet{Sweeney+22}. We only display the COs that will remain bounded to the Galaxy in \citet{Sweeney+22} for consistency.}
    \label{fig:BH_Velocity_Motion}
\end{figure*}

\citet{Sweeney+22} recently analyzed the distribution of COs, including natal kicks, from the entire Galactic population (thin disk,  thick disk, the stellar halo, and bulge). As suggested in Figure \ref{fig:3DPopulation}, the COs originating from the GC may also reach large distances. Below, we compare the GC population to the full Galactic population.

In Figure \ref{fig:Roman_Distribution}, we depict the Galactic latitude distribution of compact objects ejected from the central parsec of the GC after $3$~Gyr and within a galactocentric cylindrical radius of $8$~kpc. Nearly $70\%$ of NSs (left panels) and fast kick BHs (right panels) are located at least a degree off the Galactic plane, whereas only $20\%$ of slow kick BHs exhibit the same characteristic. Due to the strong natal kicks, the distribution of neutron stars and fast kick black holes peaks near $3\degree$ off the galactic plane. Notably, there is a subset of objects ($\sim 21$\%) expelled from the central parsec that that are  completely unbound from the Milky Way. The distribution of slow kick BHs from the central parsec is concentrated within $1\degree$. The decline beyond a few degrees is attributed to the comparatively lower velocities of natal kicks. 
We propose that the GC population can be differentiated from the rest of the Galactic population. In Figures \ref{fig:Roman_Distribution} and \ref{fig:BH_Velocity_Motion}, we compare our results to the publicly available simulation data in \citet{Sweeney+22}, note that the COs from the galactic population are $\sim 10^4$ times more numerous \footnote{Note that \citet{Sweeney+22} COs were integrated up to the present day for continuous star formation in the Milky Way, while our COs were integrated to present day from a single star formation episode in the GC 3 Gyr ago}.  As shown in Figure \ref{fig:Roman_Distribution}, the galactic population's distribution of NSs and BHs are preferentially located at higher galactic latitudes compared to the GC's population. Thus allowing for the potential differentiation of these populations.

In Figure \ref{fig:BH_Velocity_Motion}, we display the spatial and velocity distributions of the two BH populations ejected from the central parsec of the GC. As expected, the slow kick BHs are more concentrated towards the GC and remain closer to the galactic plane compared to the fast kick BHs population (see left panel). We note that the galactic population of NSs (and slow kick BHs) in \citet{Sweeney+22} extends well beyond the GC distribution in both the x and z directions. This is because the natal kicks are occurring throughout the galaxy and are not localized to the GC. 
The right panel shows that the galactic population can reach higher velocities (max. $\sim$ $870$~km~sec$^{-1}$) while the GC population attains slightly lower velocities (max. $\sim$ $730$~km~sec$^{-1}$).

\subsection{COs unbound to the Milky Way}
\label{Hypervelocity}

COs with velocities exceeding the escape velocity of the Milky Way are unbound to the Milky Way. $21\%$ of all NSs within the central parsec are unbound to the Milky Way by $3$~Gyr (average speed of $ \simeq 800$~km~sec$^{-1}$, at $100$~kpc from the center). As expected, the percentage of BHs unbound from the Milky Way depends on the underlying kick prescription. The fast kick BHs follow the NS percentage, while slow kick BHs only result in $2\%$ of BHs being unbound to the Milky Way (average speed of $ \simeq 1650$~km~sec$^{-1}$, at $100$~kpc from the center).

\subsection{Distinguishing between kick prescriptions with gravitational microlensing}
\label{RomanMicroLensing}

The different BH natal kick prescriptions predict different distributions of compact objects as a function of Galactic latitude.
Fast kicks result in an increasing number of BHs at increasing latitudes up to about $2 - 3^\circ$ off the Galactic Plane, while slow kicks result in a decreasing number of BHs at increasing latitudes (Figure \ref{fig:Roman_Distribution}).
Thus, if the number density of BHs as a function of latitude can be mapped, it would provide a way to determine the type of natal kicks BHs receive.  

Gravitational microlensing can be used to measure the masses and velocities of dark massive objects in our Galaxy; for a detailed explanation, please see \cite{Hog:1995, Miyamoto:1995, Walker:1995}.
In brief, when a foreground object (such as a BH) aligns by chance with a background star along an observer's line of sight, the gravitational field of the foreground mass deflects the background star's light.
The observer sees a transient brightening (photometric microlensing) and positional deflection (astrometric microlensing) of the background star.
These two signals can then be used to measure the mass, velocity, and distance of the unseen lens.
Gravitational microlensing has been proposed as a method to measure the mass distribution of compact objects toward the Galactic Bulge \citep{Gould:2000, Lam:2020}.

An isolated stellar-mass BH has recently been detected and characterized with microlensing, using ground-based survey photometry and Hubble Space Telescope follow-up astrometry \citep{Lam+22,Sahu+22,Mroz+22,Lam+23}.
This BH lens has been used to constrain the properties of natal kicks \citep{Andrews:2022} as well as whether the progenitor system was binary or single \citep{Vigna-Gomez:2023}.

The Nancy Grace Roman Space Telescope (Roman Space Telescope), NASA's next flagship mission scheduled to launch by 2027, will conduct several wide-field infrared surveys.
Its Galactic Bulge Time Domain Survey (GBTDS) is designed to discover thousands of cold exoplanets via gravitational microlensing \citep{Spergel:2015, Penny:2019}.
The notional design of the GBTDS\footnote{Referred to as ``WFIRST Cycle 7" in \citet{Penny:2019}.} will observe an area of $\simeq 2$ deg$^2$ around $1.5^\circ$ off the Galactic Plane, avoiding regions within a degree of the GC. 

In addition to exoplanets, the Roman Space Telescope could also detect and characterize hundreds of BHs via photometric and astrometric microlensing, as well as a comparable number of neutron stars if the astrometric precision is sufficient \citep[][although see \cite{Sajadian2023} for a more conservative estimate baed on more stringent characterization criteria]{Lam:2020,Lam:2023}. 
With its photometric precision, the Roman Space Telescope could also be used to study the population of compact objects in a statistical manner with photometric microlensing \citep{SamRose+22}. 

A detailed study is beyond the scope of this work, but we suggest that the Roman Space Telescope has the ability to study BH natal kicks and distinguish between slow and fast kicks.
In particular, including an additional pointing toward the GC in the GBTDS would enable the measurement of the BH density as a function of latitude, and enable the determination of BH kick speed.
We note that a broad range of other science cases would also be enabled by a field at the GC \citep{Terry:2023}.


\section{Discussion and Conclusion}
\label{Sec:DiscAndConclusion}

Neutron stars and perhaps even black holes receive large natal kicks during birth, with an expected average speed of $400$~km~sec$^{-1}$ \citep[e.g.,][]{Hansen+97,Arzoumanian+02,Hobbs+04}. Here we consider a GC population of massive stars (both single and binary), with different initial density distributions $\rho\sim r^{-\alpha}$, with $\alpha\in[0-3]$. The GC offers a unique opportunity to study the conditions surrounding SMBHs that probably take place in other galactic nuclei. Focusing on the post-kick density distribution and comparing it to observations allows us to infer the initial stellar distribution at our GC.

The kicks in the vicinity of the SMBH may redistribute the orbital configuration of the COs around the SMBH, as well as unbind the binary itself. Adopting a kick distribution with an average kick velocity of $400$~km~sec$^{-1}$ implies that at $\sim 0.107$~pc from the SMBH, the velocity dispersion around the SMBH is similar to that of the average kick magnitude.  Thus, overall, we expect that kicks beyond this distance will more likely be unbound COs from the SMBH (see Figure \ref{fig:schematic}), while those that remain bound \citep[based on their initial orbital configuration,][]{Lu+19}, will migrate closer to the SMBH potential.

The natal kick at the central parsec significantly affects the CO density distribution, i.e., the dark cusp. Here, we find that natal kicks steepen the resulting compact object density profiles, with most of the steepening occurring within 0.1~pc for NSs and fast BH kicks. The natal kicks are efficient at driving stellar remnants from an initial semimajor axis beyond $0.1$~pc, where the majority of the progenitor population is located, to bound orbits within $0.1$~pc from the SMBH (Figures \ref{fig:Positon} and \ref{fig:NS Density}). This result goes beyond the previous studies by \citet{Bortolas+17} and \citet{Hoang+22}, which were limited to values of 0.13pc and 0.1pc, respectively.\footnote{Note that the numerical experiment within 0.1pc, e.g., \citet{Hoang+22} yield consistent results with \citet{Bortolas+17}}. Even when considering slow black hole kicks, the resulting black hole distribution still exhibits a steepening trend, although to a lesser extent (see Appendix \ref{BH Distribution}, Figure \ref{fig: BH Density}).

Using the predicted post-natal kick CO distribution, we constrained the initial stellar profile from limits on the unseen mass within S0-2’s orbit. Specifically, observations suggested that about $\sim 4000$~M$_\odot$ reside inwards to S0-2's orbit \citep[$\lsim 1000$~au][]{Do+19,GRAVITY+20, Heibel2022}. Assuming that this unseen cusp is composed of stellar remnants such as stellar mass BHs and NSs, we infer the initial stellar density distribution. 
Considering the standard population proportions of $0.26:0.014:2.3\times 10^{-3}$ for white dwarfs, neutron stars, and black holes \citep{Kroupa01} within S0-2's orbit, an initial stellar profile with $\alpha \geq 2$  leads to a compact object density distribution that is incompatible with the mass constraints, as depicted in Figure \ref{fig:NS Density}. 

This result is consistent with current observations of the stellar density distribution as close to unity. We note that if we adopt the unseen mass to be smaller than $\sim 3000$~M$_\odot$ inwards to S0-2's orbit \citep[e.g.,][]{GRAVITY+20}, we find a stronger constrain of the initial stellar density to be $\alpha\leq 1.5$. The relation between the initial and final distribution is possible because two-body relaxation and collision effects have negligible effects on the final distribution at these stages \citep[e.g.,][]{Rose+22,Rose+23}. Also, note that some theoretical arguments suggested that the unseen mass inwards to S0-2's orbit is consistent with the existence of intermediate-mass BH \citep[e.g.,][]{Naoz+20,Generozov+20,Zhang+23,Will+23,Strokov+23}. In this case, the inferred initial stellar distribution may be even shallower. 

In addition to the steepening of the CO density profiles, natal kicks naturally lead to the creation of EMRIs, X-ray binaries, TDEs, and binary GW mergers. From these, EMRIs are the most sensitive to the initial stellar profile, with a few hundred EMRIs expected for the steepest stellar profiles, as depicted in Figure \ref{fig:Observables_at_SOI}. TDEs and binary GW mergers are less sensitive to the initial stellar profile, and we'd only expect a handful of them. The number of EMRIs and TDEs expected from natal kicks is largely negligible compared to two body relaxation processes around a single SMBH \citep[e.g.,][]{Hopman+Alexander05,Hopman+Alexander06,Alexander05,Sari+Fragione19,Fragione+Sari18}, both are much lower compared to the expectation in SMBH binaries \citep[e.g.,][]{Naoz+22,Mazzolari+22,Mockler+23,Melchor+23,Naoz+23}.  Unsurprisingly, X-ray binaries are unaffected by their distribution around the SMBH because the orbital properties of the inner binary directly affect the occurrence rate of X-ray binaries.  

A significant fraction of compact objects are unbound from the SMBH due to their natal kicks and may be potential microlensing events detectable by the the Roman Space Telescope. As a proof of concept, we follow the unbound compact objects formed from an initial distribution of $\alpha = 1.5$. This distribution is consistent with our aforementioned findings as well as with the observed GC stellar distribution \citep[e.g.,][]{Schodel+18,Gallego-Can+18}{}. We follow these COs as they migrate throughout the galaxy for 3 Gyr (see Figure \ref{fig:3DPopulation}). The adopted kick prescription is reflected in the spatial distribution of the compact objects in the galaxy. Specifically, slow-kick BHs ejected from the GC are concentrated closer toward the Galactic Plane, while fast-kick BHs and NSs are preferentially located at higher galactic latitudes.

Lastly, we compared the GC COs distribution to the expected galactic COs distribution and found that these two populations are potentially distinguishable. Particularly, the GC population is slightly slower (Fig.~\ref{fig:BH_Velocity_Motion}) and presents a longer tail towards low galactic latitude (Fig.~\ref{fig:Roman_Distribution}). The GBTDS expected field of view for the Roman Space Telescope is located in a galactic latitude range to possibly untangle the true underlying kick prescription for BHs.

\section*{Acknowledgements}
We would like to thank the anonymous referee for their feedback and efforts towards improving the manuscript.
We thank David Sweeney for useful discussions.  S.N. acknowledges the partial support from NASA ATP 80NSSC20K0505 and from NSF-AST 2206428 grant as well as thanks Howard and Astrid Preston for their generous support.
C.Y.L. acknowledges support from NASA FINESST grant No. 80NSSC21K2043 and a Carnegie Fellowship.

\appendix 

\section{CO Density Distribution}
\label{CO Density Slopes Calc}

The total number of COs at any given time is conserved because no COs are destroyed or added to the initial population. 
Therefore, 
\begin{equation}
\label{equality}
    \frac{dN_{t}(r)}{dr} = \frac{dN_{b,0}(r)}{dr} + \frac{dN_{u,0}(r)}{dr} ,
\end{equation}
where $dN_{t}(r)$ is the number of CO progenitors that are initially formed, $dN_{b,0}(r)$ is the number of bound CO progenitors before applying the effect of their natal kick, and $dN_{u,0}(r)$ is the number of CO progenitors that will be unbound due to their natal kick, all of which are in a bin of width dr at a radius $(r)$ away from the SMBH.

After the natal kicks, the COs will be scattered to different values of r and in some regions, there will be an overabundance of COs and in others a dearth. We can determine what the new slopes for the bound and unbound population will be. At a given value of r, we can compute the number of COs that now inhabit the region over the initial number of CO progenitors to determine the new slope. Dividing Equation \ref{equality} by $\frac{dN_{t}(r)}{dr}$ yields,
\begin{equation}
    1 = \frac{dN_{b}(r)}{dN_{t}(r)} + \frac{dN_{ub}(r)}{dN_{t}(r)} .
\end{equation}
In the case that $\frac{dN_{ub}(r)}{dN_{t}(r)} \ll 1$ and the fact that $dN = 4 \pi r^2 n dr$ for a spherical distribution, where n is the power-law density cusp $n = n_0 \, r^{-\alpha}$, gives 
\begin{equation}
\label{unity}
    1 = f_{b} \; r^{\alpha_t - \alpha_b} , 
\end{equation}
where $f_b = \frac{n_b(r)}{n_t(r)}$ is the relative number density between the initial population and the bound population that are in a bin of width $dr$ at a radius $r$ away from the SMBH.

Equation \ref{unity} can be rearranged to calculate the 
post-kick alpha value of the resulting CO distribution:
\begin{equation}
    \label{AlphaEquation}
    \alpha_b = \alpha_t - \frac{log(1/f_b)}{log(r)} .
\end{equation}


To determine the resulting CO density slopes, we generate a post-kick histogram distribution of COs in log space. For each bin where the fraction of unbound COs is less than $5 \%$, we apply Equation \ref{AlphaEquation} to determine the post-kick value of $\alpha$. 
The steepest initial profiles have a larger unbound fraction closer to the SMBH and provide less measurements for the value of alpha at each point. In the cases where alpha has noticeable variations, we determine the mean value for alpha to generate the slope lines in Figure \ref{fig:NS Density}.

\section{The Slow Kick Black Hole Density Distribution}
\label{BH Distribution}

\begin{figure*}[h]
    \centering
    \includegraphics[width=18cm]{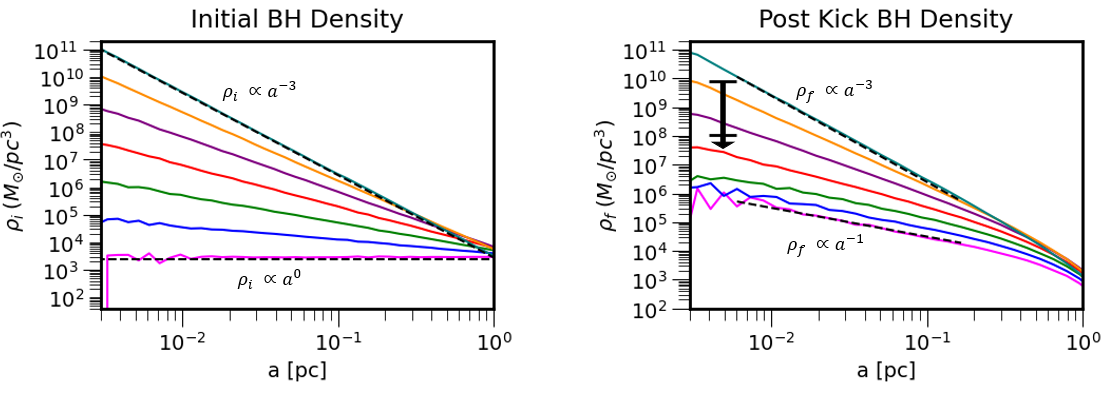}
    \caption{Density profile of BH progenitors (left) and bound BH (right) after the slow natal kicks as a function of semimajor axis. Due to the lower average kick velocity, there is not a characteristic break at $r \simeq 0.1$pc as was the case for Figure \ref{fig:NS Density}. The uppermost horizontal black line at $a = 4 \cdot 10^{-3}$ pc indicates the upper limit of the enclosed mass within S0-2's orbit with all of the mass assumed to be in BHs \citep[e.g.,][]{Heibel2022, Gravity+22}. The lower black line is the upper limit assuming a typical BH population fraction \citep{Kroupa01}. The resulting BH slopes attain a lower value of $\alpha$ than the corresponding NS Densities. }
    \label{fig: BH Density}
\end{figure*}

In Figure \ref{fig: BH Density} we show the BH progenitor and BH density distributions after the natal kicks (left panel, right panel). The post-kick slopes are estimated using the same analytical method applied to the NS distributions (see Appendix \ref{CO Density Slopes Calc}). The resulting distribution of BHs becomes steeper, with the degree of steepening being less pronounced for initially steep distributions. 

By applying a density criterion to the BH density profiles determined from the unseen mass within the orbit of S0-2, we can establish constraints on the expected initial stellar profile in the GC.\citep[e.g.,][]{Do+19, Gravity+22, Heibel2022}. The conservative upper limit is determined by assuming that the entire enclosed mass is composed of stellar-mass black holes. This limit is represented as the highest vertical black line in Figure \ref{fig: BH Density}. From this we can conclude an initial stellar profile of $\alpha < 3$ is consistent with this criteria. Note that the NS density profile provides a more stringent constraint because the resulting CO profiles are steepened due to the stronger kicks. If there are also white dwarfs and neutron stars that make up a portion of the mass fraction within S0-2s orbit, with the typical population fraction from \citet{Kroupa01}, then the upper limit is denoted by the lower vertical black line in Figure \ref{fig: BH Density}. Here an initial stellar profile with $\alpha < 2$ are allowed from the mass constraint. With a mixed population of COs, both the NS and BH density profiles converge on an upper limit, regardless of the kick distribution.  


\bibliography{Binary}
\end{document}